\documentclass[aps,twocolumn,english,superscriptaddress,citeautoscript,showkeys,preprintnumbers,amsmath,amssymb,floatfix,footinbib,prb] {revtex4-2}
\usepackage{graphicx}
\usepackage{xcolor}
\usepackage{soul}
\usepackage{amssymb,amsmath}
\usepackage{hyperref}
\hypersetup{colorlinks=true, linkcolor=red, citecolor=blue}
\usepackage[utf8]{inputenc}
\usepackage[english]{babel}
\usepackage{txfonts}
\usepackage[font=normalsize, caption=false]{subfig}
\usepackage[switch]{lineno}

\newcommand{\tc}{$T_\text{c}$}

\newcommand{\nef}{$N_{\varepsilon_\text{F}}$}
\newcommand{\omegalog}{$\omega_{\log}$}

\begin{document}


\title{First-principles evidence for conventional superconductivity in a quasicrystal approximant}

\author{Pedro N. Ferreira}
\email[Corresponding author: ]{nunesferreira@tugraz.at}
\affiliation{Institute of Theoretical and Computational Physics, Graz University of Technology, NAWI Graz, 8010, Graz, Austria}
\author{Roman Lucrezi}
\affiliation{Department of Chemistry, Stockholm University, SE-10691 Stockholm, Sweden}
\author{Sangmin Lee}
\affiliation{Department of Physics, Harvard University, Cambridge, Massachusetts 02138, USA}
\author{Lucy Nathwani}
\affiliation{Department of Physics, Harvard University, Cambridge, Massachusetts 02138, USA}
\author{Matthew Julian}
\affiliation{Enterprise Science Fund, Intellectual Ventures, Bellevue, Washington 98005, USA}
\author{Rohit P. Prasankumar}
\affiliation{Enterprise Science Fund, Intellectual Ventures, Bellevue, Washington 98005, USA}
\author{Warren E. Pickett}
\affiliation{Department of Physics and Astronomy, University of California Davis, Davis, California 95616, USA}
\author{Chris J. Pickard}
\affiliation{Department of Materials Science and Metallurgy, University of Cambridge, 27 Charles Babbage Road, Cambridge, CB3 0FS, UK}
\affiliation{Advanced Institute for Materials Research, Tohoku University, Sendai, 980-8577, Japan}
\author{Philip Kim}
\affiliation{Department of Physics, Harvard University, Cambridge, Massachusetts 02138, USA}
\author{Christoph Heil}
\email[Corresponding author: ]{christoph.heil@tugraz.at}
\affiliation{Institute of Theoretical and Computational Physics, Graz University of Technology, NAWI Graz, 8010, Graz, Austria}

\begin{abstract}

Quasicrystals (QCs) host long-range order without translational symmetry, a regime in which the very foundations of BCS theory are not straightforwardly applicable, yet experiments on QCs and their approximant crystals (ACs) point to conventional, $s$-wave, electron–phonon coupled superconductivity. Here we test the predictive power of the electron–phonon framework in a representative decagonal AC from first principles. Using state-of-the-art \textit{ab initio} methods, we compute the superconducting properties of the recently discovered AC Al$_{13}$Os$_4$ and quantitatively reproduce its bulk \tc{}. This constitutes, to our knowledge, the first \emph{ab initio} determination of \tc{} for an AC and establishes that the electron–phonon framework is predictive in these systems as well. Using the generalized quasichemical approximation for alloy modeling in the decagonal Al–Os family, we predict tunable superconductivity in Al$_{13}$Os$_{4-x}$Re$_x$ and Al$_{13}$Os$_{4-x}$Ir$_x$; in particular, Al$_{13}$Re$_4$ is dynamically stable and estimated to have a \tc{} about 30\,\% above Al$_{13}$Os$_4$. Finally, we discuss the role of ACs as high-fidelity proxies for their parent QCs. Although long-range quasiperiodicity may introduce subtle electronic features, our findings indicate that the key ingredients for superconductivity are already encoded in the local structural motifs preserved by the AC. This places the Al–Os and Al–Re families among the most promising candidates for the highest-\tc{} quasicrystalline superconductivity.
   
\end{abstract}

\date{\today}


\maketitle

\section{Introduction}

Quasicrystals (QC) represent an extraordinary class of solids that have challenged the classical laws of crystallography. They are characterized by their aperiodic yet long-range ordered atomic arrangements, exhibiting non-crystallographic rotational symmetries~\cite{levine1984, levine1986, socolar1986}.
This conceptual breakthrough was first reported in 1984 within rapidly solidified Al alloys containing 10-14\,at.\% Mn~\cite{shechtman1984}. Since then, QCs have found their way in various technological domains, including photonics~\cite{man2005, rechtsman2008, florescu2009, lin2018}, high-vacuum technologies~\cite{dubois2012}, thermal barrier coatings~\cite{dubois2012}, low-carbon steel manufacturing~\cite{liu1994}, selective laser sintering processes~\cite{kenzari2012}, catalysis~\cite{yoshimura2002,kameoka2004}, magnetism~\cite{goldman2014}, topology~\cite{chen2020, else2021}, strong correlated electrons~\cite{ishimasa2011, watanuki2012, deguchi2012}, and even in practical everyday applications, such as non-stick coatings for frying pans~\cite{dubois1993_2}.

Despite indications that QCs are not exceedingly rare in the universe~\cite{bindi2020}, superconducting QCs remain exceptionally scarce. The first indications of superconductivity (SC) in a QC was reported in 1987, observed initially in the icosahedral phase of Mg$_3$Zn$_3$Al$_2$~\cite{graebner1987}\footnote{although this phase might, in fact, be better classified as an approximant quasicrystal (AC)~\cite{takeuchi1995}, which are periodic phases resulting from QCs projection along rational directions}, in the Frank-Kasper phases of Al--Cu--Li and Al--Cu--Mg alloys~\cite{wagner1988}, and in the rapidly quenched Ti--Zr--Al alloys~\cite{azhazha2002}. 

Nevertheless, none of these initial reports provided comprehensive evidence for SC by simultaneously demonstrating zero resistivity, Meissner effect, specific heat jump at the superconducting transition, and explicit quasiperiodic symmetry within the same sample. This milestone was achieved conclusively only in 2018, with the characterization of an emergent bulk superconducting state at a critical temperature (\tc{}) of 0.05\,K in the so-called Bergmann-type Al--Zn--Mg QC alloys~\cite{kamiya2018}.

Earlier theoretical studies had predicted that Cooper pairs in structures like the Penrose lattice would exhibit exotic, unconventional superconducting pairing symmetries characterized by finite center-of-mass momentum, fundamentally driven by the lack of translational symmetry~\cite{sakai2017}. Indeed, numerous intriguing phenomena are theoretically anticipated to emerge from quasiperiodic tilings~\cite{takemori2024}, including sign-alternating superconducting pairing under magnetic fields~\cite{sakai2019}, phason-induced SC~\cite{cunyun2023}, topological SC~\cite{fulga2016, ghadimi2021, aksel2024}, and local violations of the current conservation law~\cite{fukushima2023, liu2023, fukushima2023_2}. Nevertheless, none of these predicted exotic phenomena have been experimentally observed thus far~\cite{kamiya2018, tokumoto2024, terashima2024}. In fact, contrary to more exotic mechanisms suggested theoretically, experimental measurements in both the Al--Zn--Mg QCs and their corresponding ACs indicate an $s$-wave superconducting pairing consistent with a conventional BCS weak-coupling electron-phonon (el-ph) mechanism~\cite{kamiya2018}. 

\begin{figure*}[t]
	\includegraphics[width=\linewidth]{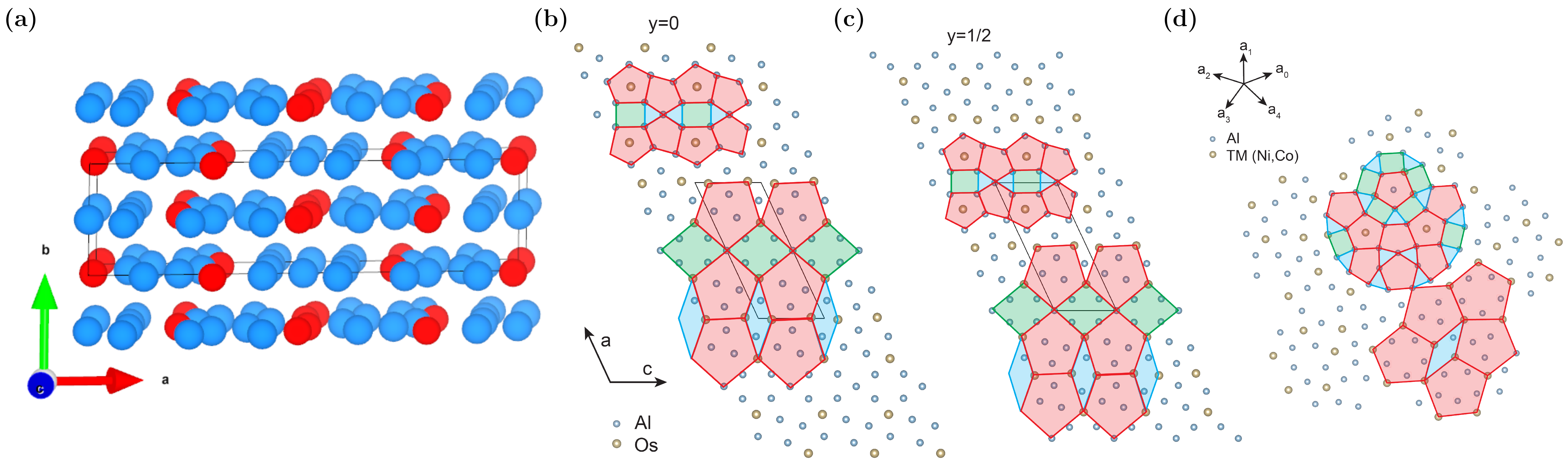}
    \caption{Crystal structure of Al$_{13}$Os$_{4}$. (a) Primitive cell (1 formula unit) of Al$_{13}$Os$_{4}$. Aluminum is represented as blue spheres and osmium as red spheres. (b) The y = 0 layer of Al$_{13}$Os$_4$ with one unit cell indicated as thin black lines. (c) The y = 1/2 layer of Al$_{13}$Os$_4$. \textbf{(d)} Variations of the smaller and larger tilings can be used in the quasicrystal Al$_{70}$Ni$_{15}$Co$_{15}$ as marked. Projections of the quasicrystal's 5 basis axes are also shown.}
	\label{fig:Al13Os4_cell}
\end{figure*}

Despite the often anticipated unconventional superconducting mechanisms in QCs, there is also support for more conventional scenarios. For instance, in Ref.~\cite{araujo2019}, the authors employed the inhomogeneous Bogoliubov–de Gennes (BdG) mean-field theory to explore SC within a two-dimensional Ammann–Beenker tiling. Predictions from this model closely align with experimental observations for the Al--Zn--Mg QC, indicating that BdG solutions for QCs can indeed reproduce conventional, BCS-like SC~\cite{araujo2019}. Furthermore, the authors in Ref.~\cite{araujo2019} provided theoretical evidence that ACs successfully capture the superconducting behavior of QCs, demonstrating that the unique electronic wavefunction characteristics of an infinite QC may not significantly influence its superconducting properties~\cite{araujo2019}. 
In fact, as QCs lack translational symmetry, ACs provide the most practical way to simulate and gain insights into their parent QCs and have therefore been investigated via \textit{ab initio} calculations to shed light on, for instance, QCs thermodynamic~\cite{yavas2023}, electronic~\cite{cain2020}, magnetic~\cite{dorini2020}, and---beginning with the present work---also their superconducting properties.

Here, we address from first principles the question of whether SC in ACs can originate from a conventional electron–phonon coupling mechanism. To this end, we perform density functional perturbation theory (DFPT) calculations on the recently reported Al$_{13}$Os$_4$ AC superconductor~\cite{meena2024}. 

Al$_{13}$Os$_4$ is a hitherto unexplored AC~\cite{edshammar1964}, isotypic to the well-known Al$_{13}$Fe$_4$~\cite{barbier1993, grin1994} and homeotypical to Al$_{13-x}$(Co$_{1-y}$Ni$_y$)$_4$~\cite{zhang1995}, associated with the decagonal phase of the Al--Os alloy~\cite{kuo1987}. In Ref.~\cite{meena2024}, SC in Al$_{13}$Os$_4$ was established via resistivity, specific heat, and magnetization measurements, as well as muon-spin rotation spectroscopy, revealing a record {\tc} of approximately 5\,K for QCs and ACs, accompanied by unambiguous BCS-like signatures. 

With this work, we show that the bulk superconducting response of Al$_{13}$Os$_{4}$ is well described within the conventional el-ph framework. This is, to our knowledge, the first \emph{ab initio} prediction of \tc{} for an AC using Migdal-Eliashberg theory, and it reproduces the experimental superconducting energy scale in a system where measurements already indicate a fully gapped, weak-coupling $s$-wave state~\cite{meena2024}. Moreover, we predict that the substitutional solid solutions Al$_{13}$Os$_{4-x}$Re$_x$ and Al$_{13}$Os$_{4-x}$Ir$_x$ provide routes to improve superconducting properties; in particular, Al$_{13}$Re$_4$ is dynamically stable and exhibits higher {\tc} and larger $\Delta_0$ compared to Al$_{13}$Os$_4$. We therefore propose that the quasicrystalline counterparts of Al$_{13}$Os$_4$ and Al$_{13}$Re$_4$ may harbor the highest \tc{} among QCs yet.

\section{Results}
\label{sec:results}

\begin{figure*}[t]
	\includegraphics[width=\linewidth]{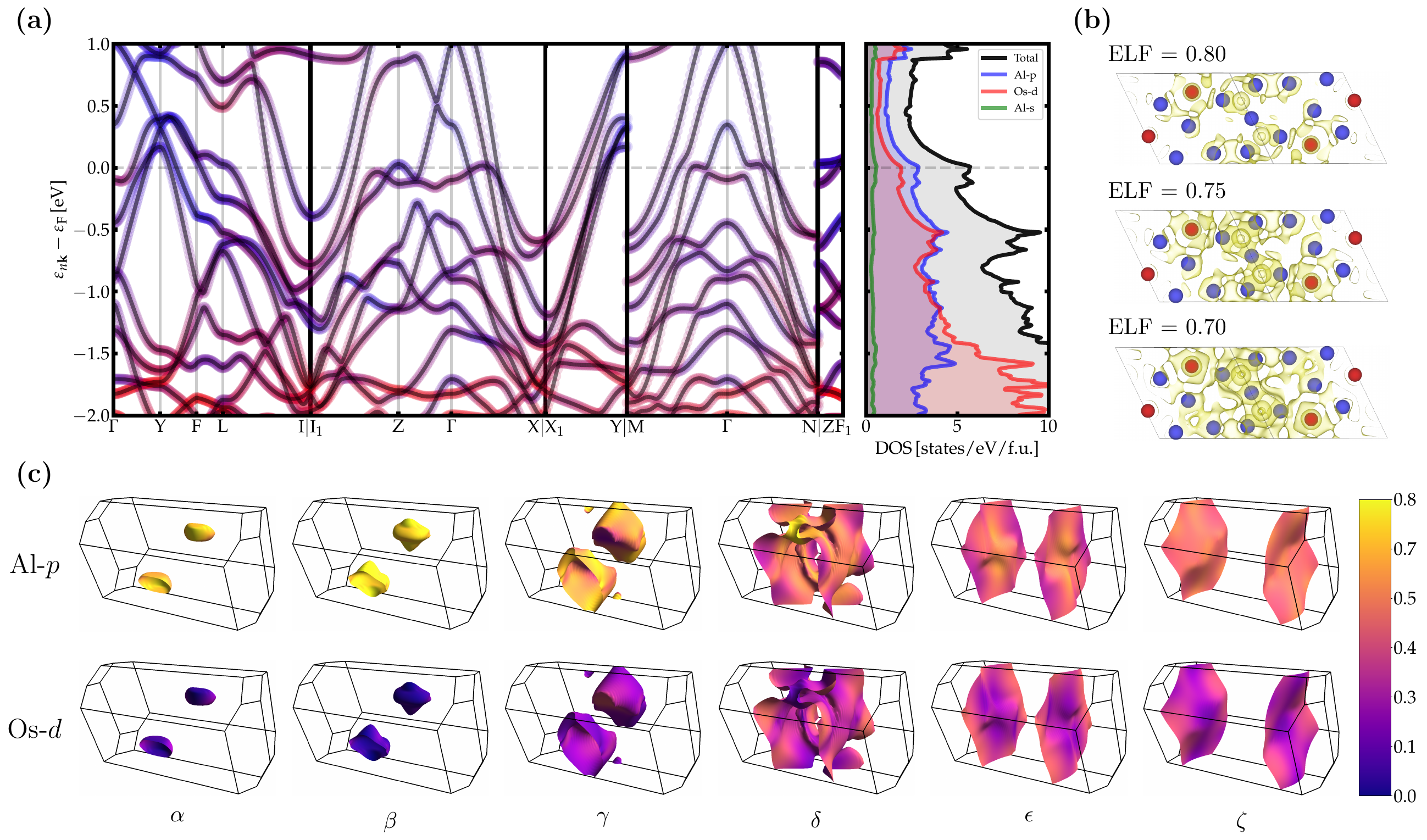}
    \caption{Electronic properties of Al$_{13}$Os$_4$. (a) Orbital-projected band structure along the selected high-symmetry path (thin black lines). Each marker’s hue is a combination of the colors assigned to the contributing orbitals, with coefficients given by their fractional projections at that state; the marker opacity encodes the total projected weight. Vertical guidelines indicate the high-symmetry points. Total electronic density of states (black, light fill) and orbital-resolved partial DOS for Al-$p$ (blue), Os-$d$ (red), and Al-$s$ (green) are also displayed. (b) Electron localization function (ELF) isosurfaces at selected values. (c) Fermi-surface sheets projected onto the Al-$p$ and Os-$d$ character.}
	\label{fig:Al13Os4_bands}
\end{figure*}

\textbf{Crystal structure.} Al$_{13}$Os$_4$ has a monoclinic structure (space group C2/m) and belongs to the family of decagonal ACs commonly designated as Al$_{13}$TM$_{4}$ (TM = transition metal)~\cite{wang1990}. It has drawn particular interest because, among known superconducting ACs, it exhibits the highest \tc{} reported so far, and its atomic arrangement closely resembles that of Al$_{13}$Fe$_4$, a compound that approximates both decagonal and icosahedral QCs~\cite{barbier1993}. Al$_{13}$Os$_4$ is isostructural to the Y–phase of the Al–Co–Ni intermetallic, Al$_{13-x}$(Co$_{1-y}$Ni$_y$)$_4$, and both of these compounds are ACs of the QC Al$_{70}$Ni$_{15}$Co$_{15}$~\cite{zhang1995,steurer1993}. Al$_{13}$Os$_4$ is also related to the metastable decagonal quasicrystal phase of the Al–Os alloy~\cite{kuo1987,wang1990}.

Fully \textit{ab initio} structural optimization of the oblique rhombic-prism primitive cell of Al$_{13}$Os$_{4}$ (17 atoms in the primitive cell, Fig.~\ref{fig:Al13Os4_cell}a) yields lattice parameters $a=9.11$\,$\AA$ and $c=7.82$\,$\AA$ and angles $\alpha=115.04^\circ$ and $\gamma=26.91^\circ$, all within 1\,\% of the experimental values. We also fully relaxed the primitive cell while accounting for spin–orbit coupling (SOC) and non-local van der Waals (vdW) interactions~\cite{berland2015,chakraborty2020}. The resulting lattice parameters differ by less than 0.2\,\% from those obtained without these effects, indicating that SOC and vdW interactions do not play a significant role in the structural properties of Al$_{13}$Os$_4$. 

\begin{figure*}[t]
	\includegraphics[width=\linewidth]{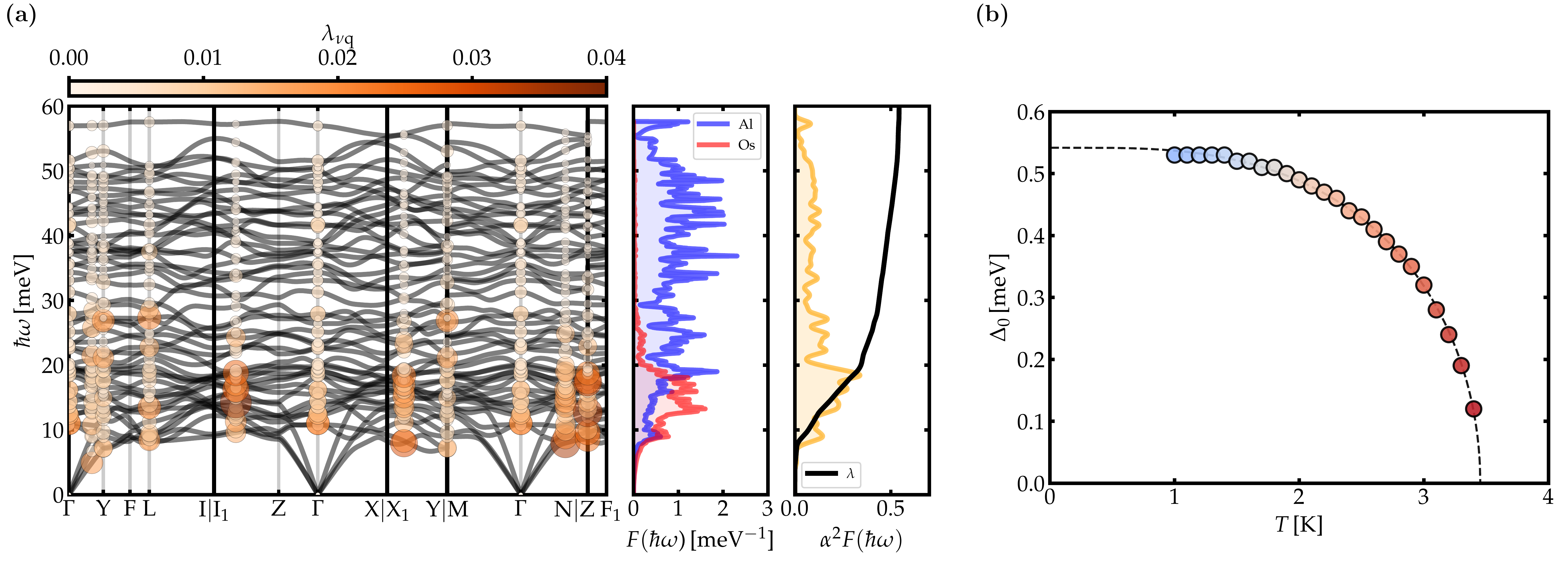}
    \caption{El–ph interactions and superconducting properties of Al$_{13}$Os$_{4}$. (a) \textit{Left:} Phonon dispersion along a high-symmetry path in the BZ. Superimposed circles indicate the mode-resolved el–ph coupling $\lambda_{\nu\textrm{q}}$ at selected $q$ points; colors and marker size encode $\lambda_{\nu\textrm{q}}$. \textit{Middle:} Atom-projected phonon density of states $F(\hbar\omega)$ in meV$^{-1}$ for Al (blue) and Os (red). \textit{Right:} Eliashberg spectral function $\alpha^2F(\hbar\omega)$ (orange line) and the cumulative total electron–phonon coupling parameter $\lambda$. (b) Superconducting gap $\Delta_0$ as a function of temperature obtained by solving the isotropic Migdal–Eliashberg equations within the full-bandwidth approximation.}
	\label{fig:Al13Os4_phonons}
\end{figure*}

Al$_{13}$Os$_4$ is layered along the \textit{b}–axis with a periodicity of $\sim$4\,\AA. The layers are structurally identical, but the layer at $y = 1/2$ is shifted by half a unit cell in the \textit{a}–axis direction (Fig.~\ref{fig:Al13Os4_cell}b and Fig.~\ref{fig:Al13Os4_cell}c)~\cite{meena2024}. Within these layers, structural motifs can be tiled in two ways: smaller tiles formed by connecting predominantly Al atoms, and larger tiles formed by connecting predominantly Os atoms~\cite{zhang1995,meena2024}. Figures~\ref{fig:Al13Os4_phonons}a,b show these tilings on each layer. The smaller tiles provide only a partial tessellation of the unit cell, whereas the larger tiles generate a gapless tiling of the entire cell. The larger tiling scheme resembles a distorted Penrose tiling composed of one pentagon and two types of rhombi: a narrow rhombus with an acute angle of $\sim$35$^\circ$, and a wider rhombus with an angle of $\sim$80$^\circ$.

Variations of both the smaller and larger tiling schemes, shown in Fig.~\ref{fig:Al13Os4_cell}d, can be used to tile Al$_{70}$Ni$_{15}$Co$_{15}$. The larger tiling in Al$_{70}$Ni$_{15}$Co$_{15}$ is a 2D Penrose tiling composed of one pentagon and one rhombus with an acute angle of $\sim$35$^\circ$. The tile shapes in Al$_{13}$Os$_4$ are slightly distorted relative to those in the ideal quasicrystal~\cite{zhang1995}.  Al$_{13}$Os$_4$ is, therefore, a promising model system for studying the emergence of quasicrystalline symmetry in Al–TM systems and for understanding the structures of other Al$_{13}$TM$_4$ ACs.


\textbf{Electrons.} To gain insights into the microscopic behavior of Al$_{13}$Os$_4$, we start with a discussion of its electronic properties. The electronic band structure along high-symmetry directions of the BZ can be appreciated in Fig.~\ref{fig:Al13Os4_bands}(a), with the states projected onto Al-$p$ (blue) and Os-$d$ (red) orbital characters.

The electronic structure of Al$_{13}$Os$_4$ is, in fact, quite complex: six distinct bands cross the Fermi level along every high-symmetry path of the BZ, as shown in Fig.~\ref{fig:Al13Os4_bands}c. The Fermi energy, $\varepsilon_{\mathrm{F}}$, lies near a local maximum in the DOS. The total DOS at the Fermi level, \nef{}, is 5.67\,states/eV/formula unit, in line with the calculated value of 5.38\,states/eV/f.u. reported in Ref.~\cite{meena2024}. Decomposing \nef{} by orbital character, 50\,\% of the states at $\varepsilon_{\mathrm{F}}$ originate from Al-$p$ orbitals and 34\,\% from Os-$d$ orbitals. The nearly equal energy dependence of the Al-$p$ and Os-$d$ DOS within $\pm 0.5$\,eV of $\varepsilon_{\mathrm{F}}$ indicates strong hybridization, corroborated by the orbital-projected bands in Fig.~\ref{fig:Al13Os4_bands}a.   

To obtain further insights into the bonding characteristics of Al$_{13}$Os$_{4}$, we computed the electron localization function (ELF); isosurfaces at selected values are shown in Fig.~\ref{fig:Al13Os4_bands}b. These isosurfaces place Al$_{13}$Os$_4$ closer to covalent metals than to simple metals: unlike simple metals, where interstitial networks typically percolate near ELF $\approx 0.5$~\cite{becke1990,desantis2000}, clear bond basins are already visible at 0.75. Consistent with this picture, two Os sites are more ionic, and the remaining two Os sites hybridize with the Al network, which partially explains the smaller Os contribution at $\varepsilon_{\mathrm{F}}$. 

We also investigated the effects of SOC and nonlocal vdW interactions on the electronic properties of Al$_{13}$Os$_4$ (see SI for details). vdW interactions produce no substantive changes. SOC lifts key degeneracies and opens a continuous pseudogap between the valence and conduction bands; nevertheless, the overall FS topography remains essentially unchanged.

Hybridization at the Fermi level can give rise to multiband behavior that may influence the superconducting properties~\cite{zehetmayer2013}. To sharpen this picture, we examined the FS topography: we resolve two small hole-like FS sheets with predominantly Al-$p$ character ($\alpha$ and $\beta$); a third hole-like sheet with pronounced Al-$p$/Os-$d$ hybridization ($\gamma$); and three additional, more extended sheets ($\delta$, $\epsilon$, $\zeta$) that exhibit strong Al–Os hybridization, as shown in Fig.~\ref{fig:Al13Os4_bands}c. 

In general, disconnected FS sheets with different electronic character lead to two or more well-separated superconducting gaps~\cite{gurevich2003, floris2007,zhao2020,correa2022,correa2023}. In Al$_{13}$Os$_4$, however, the strong hybridization among the dominant sheets implies substantial interband coupling, making a single gap on the FS plausible. The hybridized bands, additionally, are also likely to dominate the superconducting quasiparticle excitations due to their appreciable DOS, potentially suppressing any additional, smaller gap~\cite{lamura2025}. 

Consistent with this scenario, Ref.~\cite{meena2024} reports that the temperature dependence of the superconducting relaxation rate in muon spin rotation/relaxation measurements, $\sigma_{\mathrm{SC}}$, is well described by an isotropic conventional $s$-wave model. The low-temperature plateau in $\sigma_{\mathrm{SC}}$ down to the lowest measured $T$ is, indeed, consistent with a single, isotropic gap~\cite{shang2022,shiroka2022,shang2023,shang2024,panda2024,kacmarcik2025}. 

Nevertheless, there are cases where the temperature-dependent TF-$\mu$SR signal fails to distinguish single-gap to multigap order parameters~\cite{shang2025}. Hence, we believe that the single versus multigap nature of SC in Al$_{13}$Os$_4$ is not completely solved on both theoretical and experimental grounds, and further investigations along these lines would be highly valuable. A definitive theoretical resolution would require solving the fully anisotropic multiband ME equations~\cite{margine2013,lucrezi2024,mori2024} — an exceptionally demanding computation that lies beyond the actual scope of this work.

\textbf{Phonons and el-ph coupling.} Moving forward, we now examine the phonons and their coupling to electrons. We confirm that Al$_{13}$Os$_4$ is dynamically stable in the harmonic picture. The phonon dispersion and phonon density of states, shown in Fig.~\ref{fig:Al13Os4_phonons}a, indicate that the phonon spectrum extends up to 57.6\,meV, where low energy branches are hybrid Al-Os vibrations, while middle and high energy branches are due to Al.  

We also computed the el-ph coupling and found that $\lambda$ is broadly distributed across the BZ, as indicated by the size and color of the markers in Fig.~\ref{fig:Al13Os4_phonons}a. 80\,\% of the total el-ph coupling is provided by the hybrid Al-Os phonon branches up to 25\,meV. Notably, the low-frequency branches along X$_1$--Y and $\Gamma$--N host strongly coupled chiral (circularly polarized) and in-plane bending Al modes, with a smaller contribution from in-plane Os stretching. We also identify strongly coupled modes at $Z$ high-symmetry point, comprising a mixture of in-plane Al stretching and bending together with out-of-phase, out-of-plane Al/Os vibrations.  

The total el-ph coupling is $\lambda=0.54$, placing Al$_{13}$Os$_4$ in the weak-coupling regime. For comparison, the authors of Ref.~\cite{meena2024} estimated $\lambda$ from the measured \tc{} and the Debye frequency using the McMillan semi-empirical model~\cite{mcmillan1968}, obtaining $\lambda=0.63$. This independent estimate corroborates weakly coupled BCS SC and is in excellent agreement with our fully \emph{ab initio} determination. 


\textbf{Superconductivity.} To investigate the superconducting properties, we solved the isotropic Migdal–Eliashberg equations~\cite{marsiglio2020} within the full-bandwidth approximation~\cite{lucrezi2024} using the isoME code~\cite{kogler2025} to obtain a reliable estimate of \tc{} including retardation effects and the el-ph scattering processes beyond the Fermi surface. 

Within this framework, we obtain \tc{} = 3.5\,K and a zero-temperature superconducting gap $\Delta_0(0)=0.54$\,meV using $\mu^{*}_{\mathrm{AD}}=0.1$ for the Morel-Anderson Coulomb pseudopotential~\cite{morel1962}\footnote{In the IsoME code~\cite{kogler2025}, the Coulomb pseudopotential entering the Migdal–Eliashberg equations, $\mu^{*}_{\mathrm{ME}}$, is rescaled following Ref.~\cite{pellegrini2024}:
\begin{align*}
\dfrac{1}{\mu^{*}_{\mathrm{ME}}} = \dfrac{1}{\mu^{*}_{\mathrm{AD}}} + \ln\left(\dfrac{\omega_{\mathrm{ph}}}{\omega_{\mathrm{c}}}\right),
\end{align*}
where $\omega_{\mathrm{ph}}$ is the characteristic cutoff frequency for the phonon-induced interaction and $\omega_{\mathrm{c}}$ is the Matsubara frequency cutoff.}. Experimentally, Ref.~\cite{meena2024} reports onsets of the superconducting transition at 5.47\,K in resistivity, 5.45\,K in zero-field-cooled/field-cooled magnetization at 1\,mT, 5.44\,K in specific heat, and close to 5\,K in the TF-$\mu$SR relaxation rate; the magnetization offset likewise occurs near 5\,K. The experimentally fitted superconducting gap is 0.79\,meV. This represents remarkable agreement between theory and experiment for such a complex system. Indeed, the agreement is particularly compelling given the level of theory employed. Our calculations deliberately use an isotropic ME framework; achieving still higher quantitative accuracy would require a fully anisotropic treatment that accounts for the momentum and band dependence of both the el–ph and screened el–el interactions on the same footing, at substantially higher computational cost. Taken together, reproducing the correct scale of \tc{} and obtaining a gap magnitude consistent with experiment—alongside experimental indications of an isotropic $s$-wave pairing~\cite{meena2024}—show that superconductivity in Al$_{13}$Os$_4$ is quantitatively consistent with a conventional electron–phonon mechanism. While this does not uniquely exclude more complex pairing scenarios, it places strong constraints on them.

We would like to emphasize that our calculations address bulk superconductivity. Reported spin-polarized surface states in Al$_{13}$Os$_4$~\cite{meena2024} could, in principle, host additional pairing phenomena confined to the surface. Such effects would not necessarily be reflected in the bulk thermodynamic \tc{} reproduced here.

We next address an important theoretical consideration. While we have demonstrated that \textit{ab initio} calculations within the conventional el–ph framework accurately reproduce the experimental \tc{} of ACs, QCs differ in their lack of translational invariance. This absence undermines conventional prerequisites of BCS theory, including well-defined momentum space and Fermi surfaces, and in principle permits Cooper pairs with finite center-of-mass momentum. However, theoretical studies indicate that the exotic electronic wave functions of infinite QCs do not qualitatively alter their superconducting character~\cite{araujo2019}, and experimental evidence consistently points to a conventional pairing mechanism.

From a practical standpoint, standard DF(P)T cannot be directly applied to infinite QCs for \tc{} predictions due to its reliance on translational periodicity. ACs therefore serve as computationally tractable proxies that capture the essential local structural and electronic features of QCs. The key question is whether \tc{} values predicted for ACs using DF(P)T and ME theory reliably approximate those of the corresponding quasicrystalline phases. If AC-based \tc{} predictions prove reliable, they can be leveraged in first-principles calculations to identify candidate QC phases with enhanced superconducting properties.

We argue that Ref.~\cite{sakai2017}, although advocating unconventional pairing in QCs, also offers useful insights into why ACs can still provide reasonable predictions for their QCs counterparts. Using real-space dynamical mean-field theory, the authors in Ref.~\cite{sakai2017} studied the attractive Hubbard model on a Penrose lattice. 
In the weak-coupling regime, they find a finite Fourier-transformed pair amplitude of the order parameter, $\mathrm{OP}_{\mathbf{k}\mathbf{k}'}$,  beyond $\mathbf{k}'=-\mathbf{k}$ and distinct high-intensity spots along the $\mathbf{k}'=-\mathbf{k}$ line.
Nevertheless, most of $\mathrm{OP}_{\mathbf{k}\mathbf{k}'}$ remains concentrated along $\mathbf{k}'=-\mathbf{k}$, with high-intensity spots reminiscent of the BCS regime. This indicates that, even without translational invariance, a large fraction of the pair amplitude remains concentrated near $\mathbf{k}'=-\mathbf{k}$, reminiscent of the dirty-limit robustness of conventional $s$-wave pairing~\cite{anderson1959}. This motivates the expectation that the bulk \tc{} of a QC and that of a closely related AC will be comparable when their local chemistry, dominant phonon spectrum, and low-energy electronic states are similar.

In a follow-up work~\cite{takemori2020} the same authors showed that within the Penrose-tiling model, the specific-heat jump is only about 10\,\%–20\,\% smaller than the BCS value. For comparison, multiband effects in $s$-wave BCS superconductors can reduce the specific-heat anomaly by more than 20\,\%~\cite{nicol2005, nakajima2008, kacmarcik2010}, further supporting our argument.

We propose that the \tc{} of an AC serves as a physically motivated baseline for the \tc{} of its parent QC. Because the el-ph interaction is predominantly governed by the local atomic environment and short-range vibrational modes—which are essentially identical in both the AC and QC — the AC provides a reliable estimate of the pairing strength. While the long-range aperiodicity of the QC may lead to additional band-folding that could further modulate \nef{}, the AC captures the essential mechanisms of the superconducting state. In this context, the high \tc{} calculated for Al$_{13}$Os$_4$ strongly suggests that its quasicrystalline counterpart resides in a similar temperature regime. We strongly encourage experimental efforts to synthesize and characterize this material. Moreover, this strategy opens an unexplored route to predict new QCs \emph{in silico}, as we're going to demonstrate for Al$_{13}$Re${_4}$.


\textbf{Designing new superconducting ACs.} In the first report of the Al$_{13}$Os$_4$ decagonal phase in 1987~\cite{kuo1987}, the author also described the synthesis of an analogous decagonal phase in the Al–Rh system and speculated that a similar phase might exist in Al–Ir. However, to the best of our knowledge, no Al–Ir QC or their AC phases have been realized so far. Notably, iridium carries one additional $5d$ electron relative to osmium, while rhenium has one less. Thus, substitution on the transition-metal sublattice provides a way to tune the electronic properties of the system. Realizing such phases could open unexplored avenues for synthesizing high-quality, homogeneous quasicrystalline solid solutions with tunable superconducting properties, potentially surpassing those observed in the Al$_{13}$Os$_4$ AC.

Motivated by this, we investigated the thermodynamic stability and electronic properties of Al$_{13}$Os$_{4-x}$Re$_{x}$ and Al$_{13}$Os$_{4-x}$Ir$_{x}$ within the generalized quasichemical approximation (GQCA)~\cite{chen1995, ferreira2024}.

\begin{figure*}[t]
	\includegraphics[width=\linewidth]{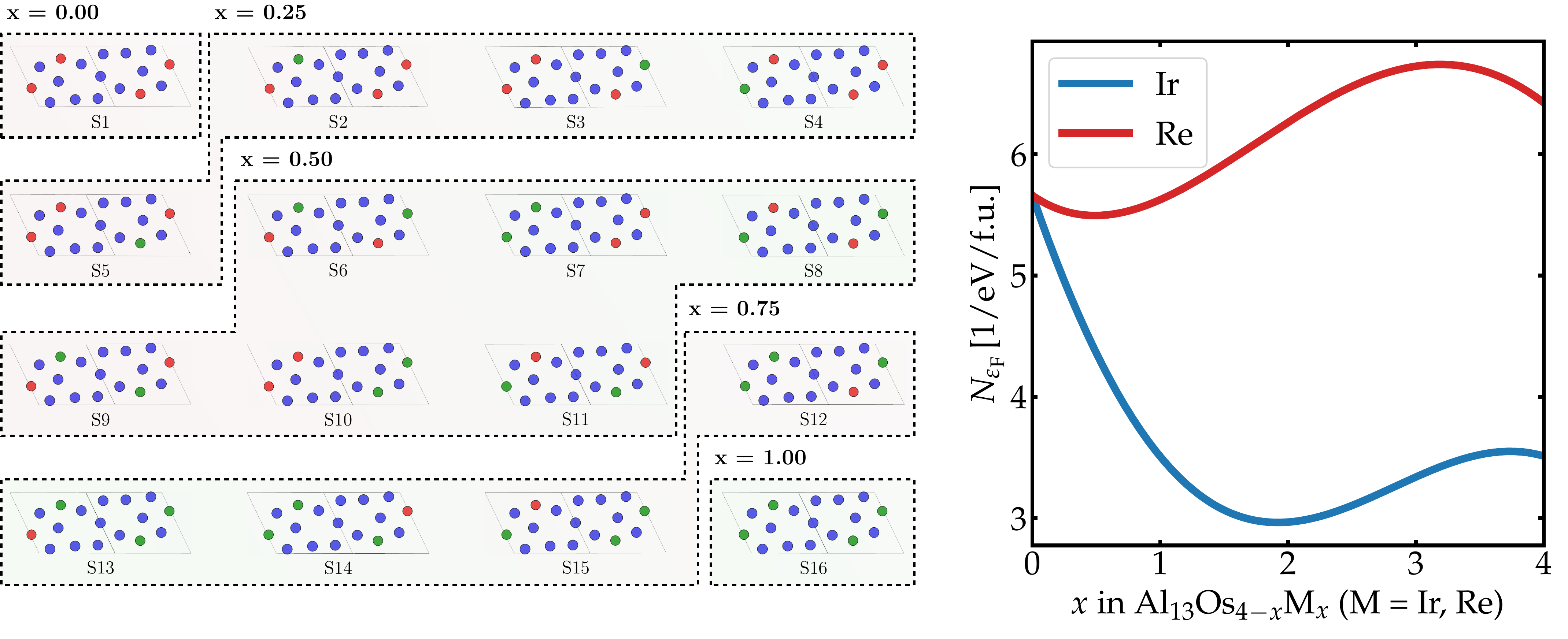}
    \caption{(left) All supercell configurations obtained by substituting Os with Re or Ir in Al$_{13}$Os$_{4}$. (right) Total electronic DOS of Al$_{13}$Os$_{4-x}$Ir$_x$ and of Al$_{13}$Os$_{4-x}$Re$_x$ as a function of composition $x$, computed within the GQCA framework.}
	\label{fig:gqca}
\end{figure*}

\textbf{Modeling Al$_{13}$Os$_4$ alloys.} To model the alloy within GQCA, we generated all possible supercell configurations by substituting osmium atoms with rhenium or iridium, resulting in 16 distinct configurations, as illustrated in Fig.~\ref{fig:gqca}(a). Among these, 13 configurations are symmetry-inequivalent. For each inequivalent configuration, we performed full \emph{ab initio} relaxations of the lattice parameters and atomic positions while constraining the $C2/m$ base-centered monoclinic Bravais lattice symmetry. The total energies of the relaxed supercells were then used to evaluate the Gibbs mixing free energy as a function of composition for Al$_{13}$Os$_{4-x}$Ir$_{x}$ and Al$_{13}$Os$_{4-x}$Re$_{x}$ from 300\,K to 1400\,K (Supplementary Fig. 2-3).

There exist clusters (supercells) with positive excess energies ($\Delta_j>0$), which yield a strongly asymmetric mixing enthalpy, $\Delta H(x,T)$, for Al$_{13}$Os$_{4-x}$Re$_x$ (Supplementary Fig.~3). This asymmetry suggests a departure from an ideal random-cluster distribution. As temperature increases, the configurational entropy dominates over enthalpy, and the Gibbs mixing free energy $\Delta G(x,T)$ becomes increasingly symmetric about $x=0.5$. Moreover, no common-tangent construction connects distinct compositions in $\Delta G$ in both systems, which, together with the positive curvature, rules out binodal and spinodal decompositions above ambient temperature. From a thermodynamic point of view, therefore, we expect that—subject to dynamical stability and kinetic constraints—high-quality, homogeneous solid solutions of Al$_{13}$Os$_{4-x}$Ir$_{x}$ and Al$_{13}$Os$_{4-x}$Re$_{x}$ should be experimentally accessible.

Now, let's examine their potential for SC. In conventional phonon-mediated superconductors, \tc{} increases with both the electron–phonon coupling parameter $\lambda$ and the logarithmic average phonon frequency \omegalog{}. This trend is captured by the modified semi-empirical McMillan expression~\cite{mcmillan1968,dynes1972,allen1975}:
\begin{align}
T_{\text{c}} = \dfrac{\omega_{\log}}{1.2} \exp \left[ - \dfrac{1.04\left( 1 + \lambda\right)}{\lambda - \mu^{*}\left( 1 + 0.62\lambda \right)} \right].
\label{eq:mcmillan}
\end{align}
As $\lambda$ is proportional to \nef{}, a well-established route to enhance \tc{} is to increase $\lambda$ by maximizing \nef{} through electronic gating or doping. However, stronger coupling is typically accompanied by a decrease of the characteristic phonon scale \omegalog{}, so our task is to identify the optimal balance between large $\lambda$ and high \omegalog{}.

With this in mind, we investigate whether \tc{} in the Al$_{13}$Os$_4$ AC can be fine-tuned by electron or hole doping via Ir or Re alloying. We then computed the total and partial DOS at $\varepsilon_F$ for the solid solutions Al$_{13}$Os$_{4-x}$Ir$_x$ and Al$_{13}$Os$_{4-x}$Re$_x$, using the GQCA ensemble average defined in Eq.~\ref{eq:gqca_average}. The results are shown in Fig.~\ref{fig:gqca}.

Ir doping suppresses the total DOS at $\varepsilon_{\mathrm{F}}$, which would be detrimental to SC. Furthermore, we found that within the harmonic approximation Al$_{13}$Ir$_4$ is dynamically unstable, exhibiting imaginary phonon modes around various high-symmetry points of the BZ (Supplementary Fig.~4). We further verified that these instabilities persist under hydrostatic pressure up to 10\,GPa. Thus, Al$_{13}$Ir$_4$ is unlikely to form, and the solid solution Al$_{13}$Os$_{4-x}$Ir$_x$ is expected to be feasible only over a narrow composition range near the Os-rich solution.

In contrast, the Al$_{13}$Os$_{4-x}$Re$_x$ solid solution provides a promising route to tune SC in Al--TM ACs. Our calculations show that \nef{} increases with Re substitution, reaching 6.57\,states/eV/f.u. in Al$_{13}$Re$_4$, where Al-$p$ and Re-$d$ states contribute nearly equally at $\varepsilon_{\mathrm{F}}$. Al$_{13}$Re$_4$ is also predicted to be dynamically stable. Since both end members are stable and the mixing thermodynamics are favorable, Al$_{13}$Os$_{4-x}$Re$_x$ is a plausible candidate for a homogeneous, continuous solid solution. In light of the enhanced \nef{}, which boosts $\lambda$, this system is expected to exhibit a higher \tc{} than pristine Al$_{13}$Os$_4$.

\begin{figure*}[t]
	\includegraphics[width=\linewidth]{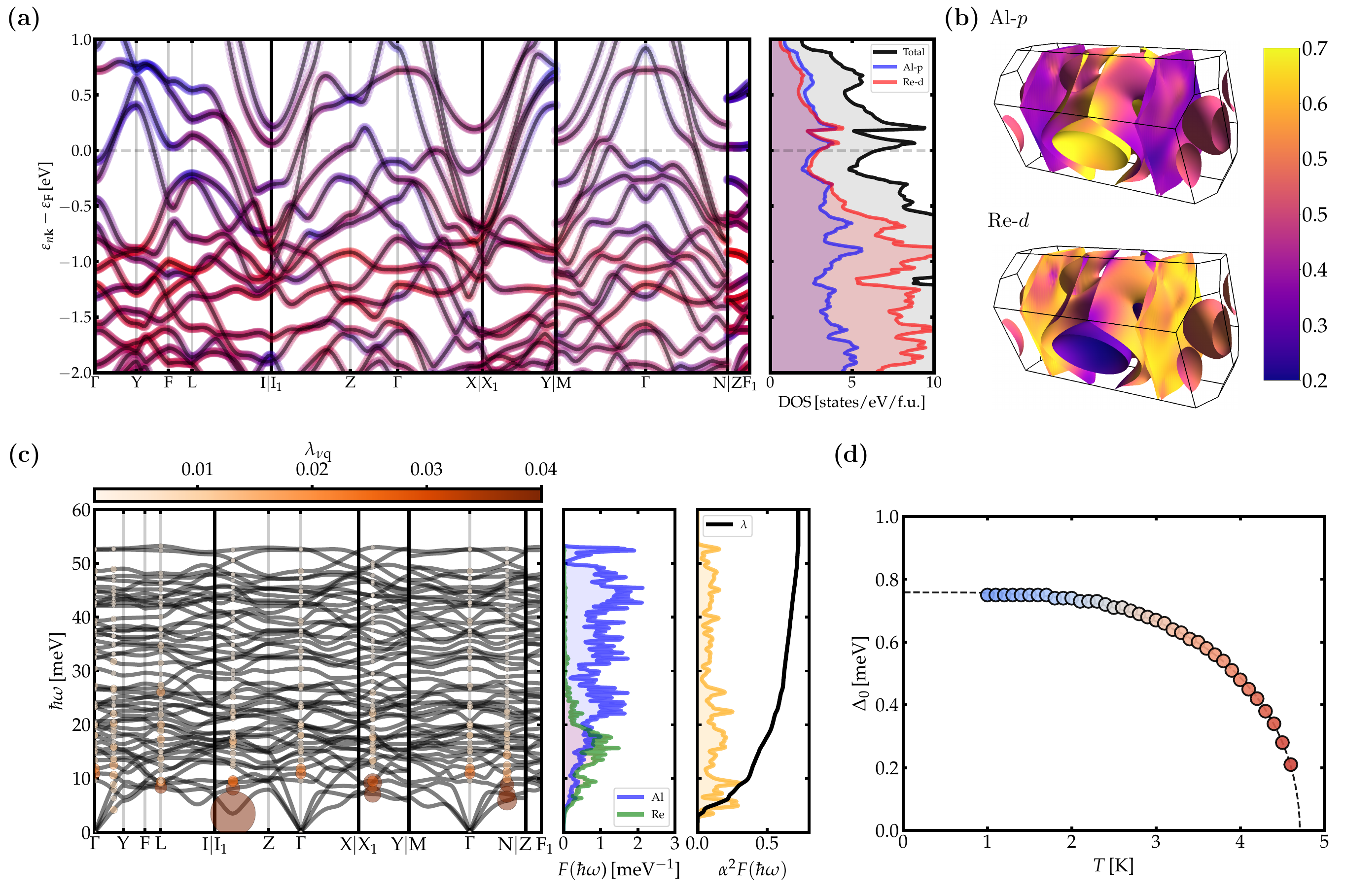}
	\caption{Electronic, el-ph and superconducting properties of Al$_{13}$Re$_4$. (a) Orbital-projected band structure, total electronic density of states (black, light fill), and orbital-resolved partial DOS for Al-$p$ (blue) and Re-$d$ (red). Each marker’s hue in the electronic dispersion is a combination of the colors assigned to the contributing orbitals, with coefficients given by their fractional projections at that state; the marker opacity encodes the total projected weight. (b) Fermi surface projected onto the Al-$p$ and Re-$d$ states. (c) \textit{Left:} Phonon dispersion along the same high-symmetry path. Superimposed circles indicate the mode-resolved el–ph coupling $\lambda_{\nu\textrm{q}}$ at selected $q$ points; colors and markers size correlate to the values of $\lambda_{\nu\textrm{q}}$. \textit{Middle:} Atom-projected phonon density of states $F(\hbar\omega)$ for Al (blue) and Re (green). \textit{Right:} Eliashberg spectral function $\alpha^2F(\hbar\omega)$ (orange line) and the cumulative total el–ph coupling parameter $\lambda$. (d) Superconducting gap $\Delta_0$ as a function of temperature obtained by solving the isotropic Migdal–Eliashberg equations within the full-bandwidth approximation.}
	\label{fig:Al13Re4}
\end{figure*}

\textbf{Superconductivity in Al$_{13}$Re$_4$}. To assess this scenario, Fig.~\ref{fig:Al13Re4} summarizes the electronic structure, phonons, el–ph coupling, and superconducting properties of Al$_{13}$Re$_4$.
In this compound, five bands cross $\varepsilon_{\mathrm{F}}$ and the FS now includes an enlarged Al-$p$–derived sheet alongside hybridized sheets.
This configuration naturally accommodates an anisotropic distribution of the superconducting gap over the FS or quasiparticle excitations with well-separated gap energies on different sheets. Consequently, Al$_{13}$Os$_{4-x}$Re$_x$ may exhibit a composition-tuned crossover from a single- (Al$_{13}$Os$_{4}$) to a multigap (Al$_{13}$Re$_{4}$) superconducting state, enriching even more the physics of these ACs and QCs and broadening the landscape of mechanisms to be explored.  

The phonon spectrum of Al$_{13}$Re$_4$ closely resembles that of Al$_{13}$Os$_4$, and $\lambda_{\mathbf{q}\nu}$ is likewise well distributed throughout the BZ, except for a single soft branch along I$_1$--Z composed of mixed in-plane and out-of-plane vibrations that couples strongly with the electrons at $\varepsilon_{\mathrm{F}}$. Since it occupies only a small fraction of $\mathbf{q}$-space, this branch does not produce a pronounced peak in the Eliashberg spectral function, and the overall electron–phonon coupling retains a distribution comparable to its Os analog. Assuming isotropic $s$-wave pairing, we predict \tc{} = 4.7\,K, about 30\,\% higher than for Al$_{13}$Os$_4$. 

Building on experimental evidence that the \tc{} of QCs closely follows that of their crystalline ACs, the Al--Re QC corresponding to Al$_{13}$Re$_4$ is therefore expected to exhibit the highest \tc{} among QCs reported to date. Notably, introducing small amounts of osmium could further enhance the \tc{} of Al$_{13}$Re$_4$ by maximizing \nef{}. 

\section{Summary}
\label{sec:summary}

In summary, our work demonstrates that fully \emph{ab initio} calculations accurately capture the superconducting state of the newly discovered Al$_{13}$Os$_4$ AC. We reproduce its bulk \tc{} and $\Delta_0$, providing, to our knowledge, the first theoretical determination of \tc{} for an AC. The close agreement with experiment offers strong evidence that the SC phase originates from a conventional el–ph coupling mechanism.

Guided by this el–ph framework, we identify Al$_{13}$Os$_{4-x}$Re$_x$ as a promising platform for enhanced superconducting properties; notably, the end member Al$_{13}$Re$_4$ is dynamically stable and is predicted to raise \tc{} by $\approx$30\,\% over Al$_{13}$Os$_4$. We further argue that \tc{} in ACs provides a practical baseline for their quasicrystalline relatives, highlighting the QCs related to Al$_{13}$Os$_4$ and Al$_{13}$Re$_4$ as leading candidates for the highest-\tc{} quasicrystals ever reported.

Together, these results demonstrate the computational feasibility of applying state-of-the-art Eliashberg theory to complex, large-unit-cell approximants. By successfully navigating the high computational cost associated with these systems, we provide a template for using ACs as predictive proxies. This work, therefore, paves the way for a systematic search for new superconducting QCs by leveraging the translational symmetry of their approximant counterparts to probe the local physics and el-ph pairing of the quasiperiodic state.

\section{Methods}
\label{sec:methods}

\textbf{DFT.} 
We performed density functional theory (DFT)~\cite{hohenberg1964,kohn1965} calculations using the Quantum \textsc{ESPRESSO} (QE) package~\cite{giannozzi2009,giannozzi2017}. Scalar-relativistic optimized norm-conserving Vanderbilt (ONCV) pseudopotentials~\cite{ONCV1,ONCV2} were taken from the \textsc{PseudoDojo} library~\cite{setten2018}. Exchange–correlation effects were treated within the generalized gradient approximation (GGA) using the Perdew–Burke–Ernzerhof (PBE) parametrization~\cite{perdew1996}. Self-consistent-field (SCF) cycles were converged to a threshold of 10$^{-10}$\,Ry, with a plane-wave kinetic-energy cutoff of 80\,Ry. Brillouin-zone sampling employed a $\Gamma$-centered Monkhorst–Pack \textbf{k}-point grid~\cite{monkhorst1976} of 8$\times$8$\times$12 and a Methfessel–Paxton smearing~\cite{MP-smearing} of 0.04\,Ry. This setup yields total-energy convergence well below 5\,meV/atom. Lattice parameters and atomic positions were fully relaxed until changes in total energy and residual forces were smaller than 10$^{-7}$\,Ry and 10$^{-6}$\,Ry/a$_0$, respectively.

\textbf{Phonons and el-ph coupling.}
Phonons and electron–phonon coupling (EPC) were computed within density-functional perturbation theory (DFPT)~\cite{DFPT}.

To improve the convergence of the superconducting properties with respect to the Gaussian smearing width and grid sizes, we introduce a correction factor to $\alpha^2F(\hbar\omega)$, as described in Ref.~\cite{bozier2025}:
\begin{align}
\alpha^2F_{\textrm{r}}(\hbar\omega) = \dfrac{N_{\varepsilon_F}^{\mathrm{tetra}}}{N_{\varepsilon_F}^{\mathrm{smear}}}\alpha^2F(\hbar\omega).
\end{align}
Here, $N_{\varepsilon_F}^{\mathrm{tetra}}$ is a high-quality, well-converged density of states at the Fermi level computed with the tetrahedron method on a fine \textbf{k}-point grid, $N_{\varepsilon_F}^{\mathrm{smear}}$ is the DOS at the Fermi level obtained with the Gaussian smearing used in the el-ph calculations, and $\alpha^2F_{\textrm{r}}(\hbar\omega)$ is the rescaled Eliashberg spectral function. We find this rescaling to be important for improving convergence of the el-ph coupling~\cite{bozier2025}.

To obtain the total el-ph coupling parameter, $\lambda$, we integrated $\alpha^2F_{r}(\hbar\omega)$ over the phonon spectrum:
\begin{align}
\lambda = 2 \int d\omega \dfrac{\alpha^2F_{\textrm{r}}(\hbar\omega)}{\omega}.
\end{align}
The electron–phonon matrix elements were computed on homogeneous 2$\times$2$\times$3 (6 irreducible $q$-points) and 3$\times$3$\times$4 (14 irreducible $q$-points) $q$-point grids, which were subsequently combined to obtain the results shown in Fig.~\ref{fig:Al13Os4_phonons} on a regular grid containing 19 irreducible $q$-points in the BZ. We verified, however, that the electron–phonon coupling and superconducting properties of Al$_{13}$Os$_4$ are already well converged even on a 2$\times$2$\times$2 uniform $\Gamma$-centered $q$-point grid with a reduced $k$-point sampling for the electronic states.

The phonon self-consistency threshold was set to 10$^{-16}$\,Ry. Electron–phonon matrix elements were integrated over a denser 24$\times$24$\times$36 \textbf{k}-grid using 50 double-delta smearing values in the range 0.001–0.05\,Ry; unless noted otherwise, the reported results correspond to 0.05\,Ry together with a phonon smearing of 0.25 THz for the \textbf{q}-grid integration.

The isotropic Migdal–Eliashberg equations within the full-bandwidth approximation~\cite{lucrezi2024} were solved using the \textsc{isoME} code~\cite{kogler2025}. The Matsubara frequency cutoff was set to 7,000\,meV. In variable DOS computations, the chemical potential was updated continuously. An energy cutoff of 5,000\,meV was applied to all quantities, except for the energy shift and $\mu$ update, where a lower cutoff of 2,000\,meV was used.

\textbf{GQCA.} 
Disorder and alloying were treated within the Generalized Quasichemical Approximation~\cite{sher1987,chen1995,ferreira2024}. 
In the GQCA formalism, an alloy is represented as an ensemble of non-equivalent supercells spanning the entire compositional range. These supercells are assumed to be both energetically and statistically independent of their local atomic environments, while remaining spatially homogeneous on a macroscopic scale. Within this approximation, the probability of occurrence of a given supercell at temperature $T$ and composition $x$ is given by
\begin{align}
p_{j} = \dfrac{g_{j}\eta^{n_{j}} e^{-\Delta_{j}/k_{\text{B}}T}}{\sum_{j=1}^{J}g_{j}\eta^{n_{j}} e^{-\Delta_{j}/k_{\text{B}}T}},
\label{eq:xj}
\end{align}
where
\begin{align}
\eta = \dfrac{xe^{\lambda_{\text{L}}/k_{\text{B}}T}}{1-x}.
\end{align}
Here, $g_j$ denotes the supercell degeneracy, $\Delta_j$ is the excess energy per permutable site relative to the end members of the solid solution, and $n_j$ is the number of dopant atoms in supercell $j$. Once the probabilities $p_j(x,T)$ are determined by solving the $n$-th order polynomial arising from free-energy minimization, any composition- and temperature-dependent property $P(x,T)$ accessible from first-principles can be obtained as the ensemble average
\begin{align}
P(x,T) = \sum_{j=1}^{J} p_j(x,T)P_j,
\label{eq:gqca_average}
\end{align}
where $P_j$ is the property associated with supercell $j$. A comprehensive description of the GQCA formalism and derivation of the equations is provided in Ref.~\cite{ferreira2024}.

Supercells were generated with the \textsc{Supercell} program~\cite{okhotnikov2016}.

\textbf{Visualization.} Fermi surfaces were visualized using \textsc{FermiSurfer}~\cite{kawamura2019}. Crystal structures were visualized with \textsc{VESTA}~\cite{momma2008}. The high-symmetry path in the first Brillouin zone was generated with \textsc{XCrysDen}~\cite{kokalj1999}.

\section*{Author Contributions}

PNF performed the DFT, DFPT, and GQCA calculations and wrote the main draft. 
RL did additional calculations.
SL and LN analyzed the QC tilings.
WEP and CJP participated in the discussions.
MJ, RPP, PK, and CH supervised this project. 
All authors participated in the discussions and revised the manuscript.

\section*{Declaration of competing interest}

The authors declare that they have no known competing financial interests or personal relationships that could have appeared to influence the work reported in this paper.

\section*{Acknowledments}

We thank Paul J. Steinhardt for insightful discussions. This work was supported by the Enterprise Science Fund of Intellectual Ventures.  Computational resources were provided by the Austria Scientific Computing clusters (ASC4 and ASC5) and the Ohio Supercomputing Center (OSC). PNF acknowledges support from the Austrian Science Fund (FWF) under project DOI 10.55776/ESP8588124. RL acknowledges the Carl Tryggers Stiftelse för Vetenskaplig Forskning (CTS 23: 2934).

\section*{Data availability}

All data used to prepare this manuscript are freely available on the Zenodo repository \cite{zenodo}. \\


\bibliographystyle{apsrev4-2}
\bibliography{refs}

\end{document}